\documentclass[conference,a4paper]{IEEEtran}

\ifCLASSINFOpdf
\else
\fi
\usepackage[left=1.57cm,right=1.57cm,top=0.95cm,bottom=2.54cm]{geometry}
\usepackage{listings}
\usepackage{setspace}
\usepackage{graphicx}
\usepackage{tabularx,booktabs}
\usepackage{dingbat}
\usepackage{diagbox}
\usepackage{multirow} 
\usepackage[hyphens]{url}
\usepackage[hidelinks,breaklinks]{hyperref}
\usepackage{xcolor}
\usepackage{amsfonts, amsmath, amsthm, amssymb}
\usepackage{subcaption}
\hypersetup{breaklinks=true}
\usepackage{tikz}
\usepackage{textcomp}
\usepackage{lipsum}
\usepackage{svg}
\IEEEoverridecommandlockouts
\newcommand\copyrighttext{%
  \footnotesize \textcopyright 2026 IEEE.  Personal use of this material is permitted.  Permission from IEEE must be obtained for all other uses, in any current or future media, including reprinting/republishing this material for advertising or promotional purposes, creating new collective works, for resale or redistribution to servers or lists, or reuse of any copyrighted component of this work in other works.}
\newcommand\copyrightnotice{%
\begin{tikzpicture}[remember picture,overlay]
\node[anchor=south,yshift=10pt] at (current page.south) {\fbox{\parbox{\dimexpr\textwidth-\fboxsep-\fboxrule\relax}{\copyrighttext}}};
\end{tikzpicture}%
}
\usepackage{graphicx}
\graphicspath{{figures/}}
\begin{document}

%
\title{Evaluation of NVENC Split-Frame Encoding (SFE) for UHD Video Transcoding}

\author{
    \IEEEauthorblockN{Kasidis Arunruangsirilert\IEEEauthorrefmark{1}, Jiro Katto\IEEEauthorrefmark{1}}
    \IEEEauthorblockA{\IEEEauthorrefmark{1}Department of Computer Science and Communications Engineering, Waseda University, Tokyo, Japan
    \\\{kasidis, katto\}@katto.comm.waseda.ac.jp}
}
%

\maketitle

\copyrightnotice
\setstretch{0.98}
\begin{abstract}
NVIDIA Encoder (NVENC) features in modern NVIDIA GPUs, offer significant advantages over software encoders by providing comparable Rate-Distortion (RD) performance while consuming considerably less power. The increasing capability of consumer devices to capture footage in Ultra High-Definition (UHD) at 4K and 8K resolutions necessitates high-performance video transcoders for internet-based delivery. To address this demand, NVIDIA introduced Split-Frame Encoding (SFE), a technique that leverages multiple on-die NVENC chips available in high-end GPUs. SFE splits a single UHD frame for parallel encoding across these physical encoders and subsequently stitches the results, which significantly improves encoding throughput. However, this approach is known to incur an RD performance penalty. The widespread adoption of NVIDIA GPUs in data centers, driven by the rise of Generative AI, means NVENC is poised to play a critical role in transcoding UHD video. To better understand the performance-efficiency tradeoff of SFE, this paper evaluates SFE's impact on RD performance, encoding throughput, power consumption, and end-to-end latency using standardized test sequences. The results show that for real-time applications, SFE nearly doubles encoding throughput with a negligible RD performance penalty, which enables the use of higher-quality presets for 4K and makes real-time 8K encoding feasible, effectively offsetting the minor RD penalty. Moreover, SFE adds no latency at 4K and can reduce it at 8K, positioning it as a key enabler for high-throughput, real-time UHD transcoding.

\end{abstract}



\setstretch{0.936}

%
\IEEEpeerreviewmaketitle

\vspace{-3mm}
\section{Introduction}

While the capture, encoding, and transmission of Ultra High-Definition (UHD) footage in professional production environments were developed as early as the 2010s \cite{5959148, 6775954, 7037246}, these settings typically feature fixed channel bandwidth and predictable encoding throughput demands. More recently, UHD 4K and 8K capture capabilities have been integrated into a wide range of consumer and embedded devices, including smartphones, camcorders, and cameras in autonomous vehicles and smart city surveillance systems, leading to sharp increase in UHD User Generated Content (UGC) designated for online social media platform \cite{10248488}. Concurrently, emerging Generative AI (GenAI) models such as Gemini 2.5 Pro and Veo 3 are becoming a new source of UHD video processing and generation \cite{google_inc_2025, hume_2025}, requiring UHD video payloads to be transmitted between AI data centers and end-users. In addition, new delivery mechanisms are being deployed, such as using 5G Multicast-Broadcast Services (MBS) and Wireless IP Multicasting Services in Local 5G to deliver UHD TV to residences with aging coaxial infrastructure. This approach requires encoding at the edge of the AI Radio Access Network (AI-RAN) with instantaneous throughput and bitrates varying according to household demand \cite{10913885}. As the production and consumption of UHD content by both humans and machines rapidly increases, the demand for video encoders that offer both high throughput and high Rate-Distortion (RD) performance has increased significantly.


\begin{figure}[t!]
\centering\includesvg[width=0.9\linewidth,inkscapelatex=false]{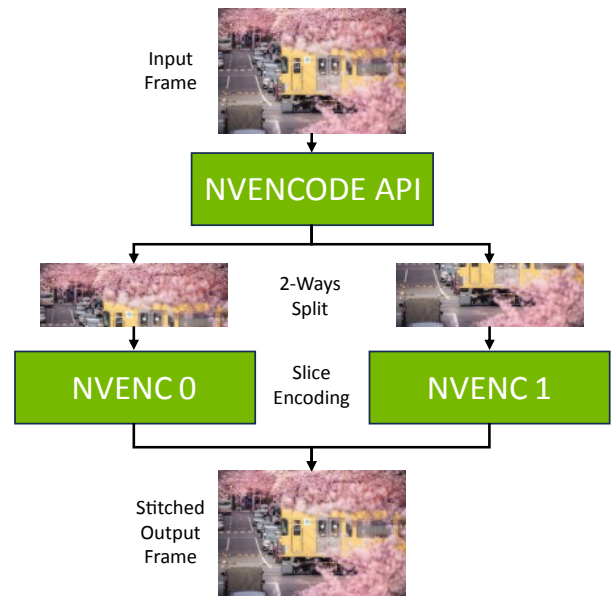}

\caption{Pipeline of NVIDIA Split Frame Encoding (SFE)}
\label{fig:Pipeline}
\vspace{-7mm}
\end{figure}

The high demand for machine learning and GenAI has driven the transformation of conventional data centers into AI-centric infrastructures through the large-scale deployment of specialized accelerators like Graphics Processing Units (GPUs) and Tensor Processing Units (TPUs) \cite{10689061, 10415538}. NVIDIA is a key manufacturer in this space, holding a dominant 92\% market share in data center GPUs as of early 2024, largely due to the popularity of its high-performance products such as the A100 and H100 \cite{joaquin_2025}. Similar to their consumer counterparts, these data center GPUs feature the NVIDIA Encoder (NVENC), a hardware-based solution that supports recent video codecs including H.264/AVC, H.265/HEVC, and AV1. NVENC provides significantly higher throughput than software-based encoders \cite{nvidia_2013}, and recent studies have shown that it can deliver comparable Rate-Distortion (RD) performance while being substantially more energy-efficient \cite{10637525}. \looseness=-1


A well-established technique for increasing the throughput of hardware video encoders involves splitting a high-resolution video frame into several slices and processing each slice on an independent encoder before stitching the outputs together \cite{6775906}. As high-end NVIDIA GPUs typically contain more than one NVENC chip \cite{nvidia_2025}, NVIDIA introduced a similar approach called \textit{Split-Frame Encoding (SFE)} in early 2024 with its Ada Lovelace architecture, supporting both H.265/HEVC and AV1 codecs \cite{nvidia_2024}. By dividing a UHD frame into multiple slices and assigning each to a dedicated NVENC chip for parallel processing (see Fig. \ref{fig:Pipeline}), SFE significantly improves encoding throughput, enabling real-time encoding of 8K video at 60 frames per second. However, this method is known to incur an RD performance penalty because each encoder operates independently, which limits the ability to exploit spatial correlations across slice boundaries. While NVIDIA suggests this penalty is minor, not exceeding a 4\% degradation in PSNR, their analysis used bitrates significantly higher than those typical for online transcoding, which may obscure the true performance impact. Furthermore, the use of proprietary video content in their evaluation prevents direct and fair comparisons to other solutions.


Since NVENC is an integrated component of NVIDIA GPUs, its deployment has become widespread in large data centers, edge servers, and AI-RAN as a direct result of the global buildout for AI workloads. Consequently, NVENC is positioned to become a dominant technology for video transcoding, with SFE expected to be a key enabler for accelerating UHD tasks. Understanding the performance tradeoffs associated with enabling SFE is paramount to designing the next generation of video delivery systems. Therefore, in this paper, NVIDIA SFE is evaluated against non-SFE configurations in terms of RD performance, encoding throughput, power consumption, and end-to-end latency. The analysis is conducted across various presets and tuning combinations using the standardized ITE's Ultra-high definition/wide-color-gamut standard test sequences (Series A) \cite{ITE_2016}, which allows for a fair comparison to other encoder and accelerator solutions.\looseness=-1

\vspace{-1mm}
\section{Experiment Setup}

\subsection{Performance Evaluation}

\begin{table}[!tbp]
\setstretch{0.85}
\caption{Target Encoding Bitrates for each resolution}
\vspace{-1.5mm}
\centering
\label{tab:bitrateRange}
\resizebox{7.5cm}{!}{\begin{tabular}{@{}lc@{}}
\toprule
Resolution                 & Bitrates (Mbps) \\\midrule
3840x2160 (2160p/4K) & 10, 15, 20, 25, 30, 35, 40, 45, 50\\
7680x4320 (4320p/8K) & 20, 30, 40, 50, 60, 70, 80, 90, 100\\
\bottomrule
\end{tabular}}
\vspace{-6mm}
\end{table}


The evaluation was performed using The Institute of Image Information and Television Engineers (ITE) Ultra-high definition/wide-color-gamut standard test sequences (Series A). This dataset includes eleven 8K and ten 4K sequences at 59.94 frames per second, with resolutions conforming to the ITU-R standard. To reflect the characteristics of typical internet-delivered content, the source sequences, provided in 12-bit BT.2020 (HDR) DPX format, were converted to 8-bit Standard Dynamic Range (SDR) with 4:2:0 chroma subsampling using the standardized method provided by ITE. To create a high-quality source compatible with hardware decoders, this SDR dataset was then encoded into an All-Intra H.265/HEVC format using the \textit{x265} encoder with medium preset at a Constant Rate Factor (CRF) of 10. All VMAF and PSNR calculations in this study were performed against this H.265/HEVC encoded reference, not the original DPX files. The evaluation was conducted on a NVIDIA GeForce RTX 4070 Ti SUPER, which shares the Ada Lovelace microarchitecture with data center-grade GPUs like the NVIDIA L40. As demonstrated in prior work \cite{10637525}, NVENC performance is consistent across GPUs of the same generation, making these results representative of typical data center deployments. VMAF was chosen as the primary metric due to its widespread industry adoption for perceptual quality assessment, though it is acknowledged that its 4K-centric model is a potential limitation for 8K evaluation. \looseness=-1

\begin{table}[!tbp]
\setstretch{0.85}
\caption{Hardware and Software Configuration}
\vspace{-1.5mm}
\centering
\label{tab:hardware}
\resizebox{8.2cm}{!}{\begin{tabular}{@{}ll@{}}
\toprule
\multicolumn{2}{c}{Encoding System}\\
\midrule
Hardware                 & Description  \\\midrule
CPU & AMD Ryzen 9 5900X 12-Core Processor \\
RAM & Dual-Channel DDR4 96 GB @ 3600 MHz \\ 
GPU & NVIDIA GeForce RTX 4070 Ti SUPER\\\midrule
Software & Version \\\midrule
OS & Microsoft Windows 10 Pro Build 19045\\
ffmpeg & 2025-06-28-git-cfd1f81e7d-full\_build-www.gyan.dev\\
NVIDIA GPU Driver & GeForce Game Ready Driver 572.42\\
Open Broadcaster Software & 31.1.1\\
Simple Realtime Server (SRS) & ossrs/srs:6.0.166 \\\midrule
\multicolumn{2}{c}{VMAF Calculation System}\\
\midrule
Hardware                 & Description  \\\midrule
CPU & AMD Ryzen Threadripper 3960X 24-Core Processor \\
RAM & Quad-Channel DDR4 128 GB @ 3200 MHz \\ 
GPU & 2x NVIDIA GeForce RTX 3090\\\midrule
Software & Version \\\midrule
OS & Ubuntu 22.04.5 LTS\\
ffmpeg & N-120252-g3ce348063c\\
libvmaf & v3.0.0 (b9ac69e6)\\
VMAF Model & vmaf\_4k\_v0.6.1neg \\
NVIDIA GPU Driver & 575.64.03 \\
NVIDIA CUDA Compiler & cuda\_12.9.r12.9\/compiler.36037853\_0 \\
\bottomrule
\end{tabular}}
\vspace{-3mm}
\end{table}


To select appropriate bitrate ranges for the evaluation, a preliminary study was conducted. The ITE dataset was uploaded to YouTube to analyze the bitrates used by a typical online video platform. The resulting 4K VP9 encodes ranged from 9.46 to 50.9 Mbps, averaging 29.5 Mbps. For 8K content, YouTube used the AV1 codec with bitrates from 11.8 to 63.6 Mbps, averaging 40.0 Mbps. These figures were considered alongside the standards for commercial UHD TV services in Japan, which deliver 4K and 8K content via H.265/HEVC at approximately 25 Mbps and 95 Mbps, respectively. Based on this analysis, nine target bitrates were chosen for each resolution, as detailed in Table \ref{tab:bitrateRange}.

\begin{table}[!tbp]
\setstretch{0.85}
\caption{Encoding Configurations}
\vspace{-1.5mm}
\centering
\label{tab:tuning}
\resizebox{7.5cm}{!}{\begin{tabular}{@{}lccc@{}}
\toprule
Test Case & tune & 2pass & multipass  \\\midrule
Ultra High-Quality (2 Pass) & 5 & 1 & 2 (Full Resolution)\\
Ultra High-Quality & 5 & 0 & 0\\
High-Quality & 1 & 0 & 0\\
Low-Latency & 2 &  0 & 0\\
Ultra Low-Latency & 3 & 0 & 0\\
\bottomrule
\end{tabular}}
\vspace{-6mm}
\end{table}


The test sequences were encoded using the system described in Table \ref{tab:hardware}. Both H.265/HEVC and AV1 codecs were evaluated across five tuning configurations (Table \ref{tab:tuning}) and three presets: P1 (Fastest), P4 (Medium), and P7 (High-Quality). Since the GPU used in the test system, the NVIDIA GeForce RTX 4070 Ti SUPER, contains two NVENC chips, only two-way SFE was supported and evaluated in this work. Following the recommendations from online streaming platforms such as YouTube and Twitch \cite{google, twitch}, a Group of Pictures (GOP) size of 120 frames (two seconds) and Constant Bitrate (CBR) rate control were used. SFE was enabled or disabled by setting the \textit{{split\_encode\_mode}} parameter to 2 or 15, respectively. Additionally, both spatial and temporal Adaptive Quantization were disabled. It should be noted that \textit{b\_ref\_mode} and \textit{b\_adapt} were only active for the tuning that enables B-frame (HQ, UHQ, and UHQ 2 Pass). All encodings were performed using the following command:

\begin{lstlisting}
ffmpeg -c:v hevc_cuvid -i {input} -vcodec {hevc_nvenc/av1_nvenc} -b_ref_mode 1 -b_adapt 1 -preset {preset} -rc cbr -tune {tune} -spatial-aq 0 -temporal-aq 0 -split_encode_mode {split_encode_mode} -2pass {2pass} -multipass {multipass} -rc-lookahead 0 -an -b:v {target_bitrate} -g 120 {output}
\end{lstlisting}



\begin{table*}[!tbp]
\setstretch{0.83}
\caption{Average Peak Signal-to-Noise Ratio (PSNR) and VMAF scores for configurations with and without Split-Frame Encoding (SFE). The scores are averaged across all test sequences and target bitrates for each combination of resolution, codec, preset, and tuning. The `Diff.' column quantifies the performance change when SFE is enabled; a negative value indicates a degradation in quality.}
\vspace{-1.5mm}
\centering
\label{tab:RDResults}
\resizebox{16cm}{!}{\begin{tabular}{@{}lc ccc ccc ccc ccc@{}}
\toprule 
\multicolumn{14}{c}{UHD 4K (2160p)}\\
\midrule
\multirow{4}{*}{Tuning} & \multirow{4}{*}{Preset} & \multicolumn{6}{c}{H.265/HEVC} & \multicolumn{6}{c}{AV1}\\ 
\cmidrule(lr){3-8} \cmidrule(lr){9-14} & & \multicolumn{3}{c}{PSNR (dB)} & \multicolumn{3}{c}{VMAF (4K)} & \multicolumn{3}{c}{PSNR (dB)} & \multicolumn{3}{c}{VMAF (4K)} \\
\cmidrule(lr){3-5}\cmidrule(lr){6-8}\cmidrule(lr){9-11}\cmidrule(lr){12-14} & & No SFE & SFE & Diff.& No SFE & SFE & Diff.& No SFE & SFE & Diff.& No SFE & SFE & Diff.\\
\midrule
\multirow{3}{*}{Ultra High-Quality (2 Pass)} & P7 & 36.427&36.276&\textbf{-0.151}&86.219&85.830&\textbf{-0.389}&36.699&36.482&\textbf{-0.216}&87.535&86.919&\textbf{-0.615} \\
&P4&36.381&36.374&-0.007&86.037&86.015&-0.021&36.692&36.682&-0.010&87.498&87.472&-0.026 \\
&P1&36.342&35.932&\textbf{-0.410}&85.943&85.591&\textbf{-0.352}&36.589&36.346&\textbf{-0.243}&87.215&86.533&\textbf{-0.681} \\
\midrule
\multirow{3}{*}{Ultra High-Quality}&P7&36.390&36.387&-0.003&86.104&86.086&-0.018&36.733&36.724&-0.009&87.622&87.612&-0.010 \\
&P4&36.336&36.332&-0.004&85.887&85.869&-0.017&36.721&36.711&-0.009&87.559&87.545&-0.014 \\
&P1&36.297&36.294&-0.004&85.793&85.780&-0.013&36.617&36.608&-0.009&87.283&87.270&-0.013 \\
\midrule
\multirow{3}{*}{High-Quality}&P7&36.266&36.257&-0.010&85.804&85.786&-0.017&36.529&36.520&-0.009&87.043&87.052&0.008  \\
&P4&36.220&36.209&-0.011&85.659&85.638&-0.021&36.511&36.502&-0.009&86.967&86.975&0.008  \\
&P1&35.936&35.929&-0.007&85.604&85.607&0.004&36.389&36.382&-0.008&86.654&86.667&0.012   \\
\midrule
\multirow{3}{*}{Low-Latency}&P7&36.097&36.094&-0.003&86.148&86.158&0.010&36.328&36.317&-0.011&86.673&86.667&-0.006  \\
&P4&36.089&36.084&-0.005&86.101&86.114&0.013&36.280&36.267&-0.013&86.600&86.588&-0.012  \\
&P1&35.936&35.929&-0.007&85.604&85.607&0.004&36.072&36.055&-0.017&86.243&86.212&-0.031  \\
\midrule
\multirow{3}{*}{Ultra Low-Latency}&P7&36.097&36.094&-0.003&86.148&86.158&0.010&36.328&36.319&-0.009&86.673&86.673&0.000   \\
&P4&36.089&36.084&-0.005&86.101&86.114&0.013&36.280&36.267&-0.013&86.600&86.588&-0.012  \\
&P1&35.936&35.929&-0.007&85.604&85.607&0.004&36.072&36.055&-0.017&86.243&86.212&-0.031  \\
\midrule
\multicolumn{14}{c}{UHD 8K (4320p)}\\
\midrule
\multirow{4}{*}{Tuning} & \multirow{4}{*}{Preset} & \multicolumn{6}{c}{H.265/HEVC} & \multicolumn{6}{c}{AV1}\\ 
\cmidrule(lr){3-8} \cmidrule(lr){9-14} & & \multicolumn{3}{c}{PSNR (dB)} & \multicolumn{3}{c}{VMAF (4K)} & \multicolumn{3}{c}{PSNR (dB)} & \multicolumn{3}{c}{VMAF (4K)} \\
\cmidrule(lr){3-5}\cmidrule(lr){6-8}\cmidrule(lr){9-11}\cmidrule(lr){12-14} & & No SFE & SFE & Diff.& No SFE & SFE & Diff.& No SFE & SFE & Diff.& No SFE & SFE & Diff.\\
\midrule
\multirow{3}{*}{Ultra High-Quality (2 Pass)}&P7&38.174&38.167&-0.007&85.104&85.096&-0.008&38.408&38.396&-0.012&86.789&86.757&-0.032  \\
&P4&38.132&38.123&-0.008&84.978&84.966&-0.012&38.533&38.522&-0.011&86.719&86.685&-0.034  \\
&P1&38.097&38.089&-0.008&84.920&84.912&-0.007&38.555&38.545&-0.011&86.391&86.353&-0.038  \\
\midrule
\multirow{3}{*}{Ultra High-Quality}&P7&38.104&38.112&0.008&84.902&84.911&0.009&38.591&38.587&-0.004&86.850&86.838&-0.012   \\
&P4&38.058&38.067&0.009&84.767&84.781&0.014&38.571&38.567&-0.003&86.774&86.761&-0.012   \\
&P1&38.010&38.022&0.012&84.679&84.711&0.032&38.435&38.431&-0.004&86.433&86.422&-0.011   \\
\midrule
\multirow{3}{*}{High-Quality}&P7&38.002&37.968&-0.033&84.860&84.727&\textbf{-0.133}&38.369&38.362&-0.007&86.330&86.319&-0.011 \\
&P4&37.951&37.923&-0.028&84.704&84.594&\textbf{-0.110}&38.344&38.337&-0.007&86.235&86.226&-0.009 \\
&P1&37.531&37.541&0.010&84.448&84.501&0.052&38.190&38.186&-0.004&85.849&85.847&-0.003   \\
\midrule
\multirow{3}{*}{Low-Latency}&P7&37.714&37.729&0.015&85.001&85.075&0.075&38.094&38.087&-0.007&85.834&85.812&-0.022   \\
&P4&37.701&37.713&0.012&84.966&85.025&0.060&38.025&38.024&-0.001&85.732&85.725&-0.007   \\
&P1&37.531&37.541&0.010&84.448&84.501&0.052&37.794&37.788&-0.006&85.302&85.288&-0.014   \\
\midrule
\multirow{3}{*}{Ultra Low-Latency}&P7&37.714&37.729&0.015&85.001&85.075&0.075&38.094&38.087&-0.007&85.834&85.812&-0.022   \\
&P4&37.701&37.713&0.012&84.966&85.025&0.060&38.025&38.024&-0.001&85.732&85.725&-0.007   \\
&P1&37.531&37.541&0.010&84.448&84.501&0.052&37.794&37.788&-0.006&85.302&85.288&-0.014   \\
\bottomrule
\end{tabular}}
\vspace{-5mm}
\end{table*}

The VMAF score and PSNR of each encoded video were calculated against the H.265/HEVC reference file using \textit{libvmaf\_cuda}. To ensure a fair evaluation of codec performance by disregarding potential enhancement gains, the \textit{vmaf\_4k\_v0.6.1neg} model was utilized for all calculations.

To measure encoding throughput without being limited by the hardware decoder, which was identified as a potential bottleneck, black frames were generated on-the-fly using an ffmpeg filter as the input source. The target bitrate was set to 25 Mbps for 4K resolution, and 80 Mbps for 8K resolution. The total time to encode 1000 frames for 4K resolution and 500 frames for 8K resolution was measured using ffmpeg's internal timer. The command used was as follows:

\begin{lstlisting}
ffmpeg -t {40/20} -f lavfi -i color=c=black:s={3840x2160/7680x4320} -vcodec {hevc_nvenc/av1_nvenc} -preset 15 -b_ref_mode 1 -b_adapt 1 -rc cbr -tune {tune} -spatial-aq 0 -temporal-aq 0 {split_encode_mode} -2pass {2pass} -multipass {multipass} -rc-lookahead 0 -an -b:v {25M/80M} -g 120 -y {output}
\end{lstlisting}

\subsection{End-to-End Latency Evaluation}


To evaluate end-to-end latency, Open Broadcaster Software (OBS) and an SRS media server were run on the encoding system (Table \ref{tab:hardware}). A source video displaying a 60 fps timecode was streamed from OBS to the local SRS server hosted on the encoding machine via the WebRTC-HTTP Ingestion Protocol (WHIP), with encoding parameters set for each test scenario. Similar to the encoding throughput benchmark, the bitrate of 25 Mbps and 80 Mbps were used for 4K and 8K live encoding, respectively. A separate playback client on the same Gigabit Ethernet network received the stream using the WebRTC-HTTP Egress Protocol (WHEP) via a web-based player. Latency was measured by capturing a side-by-side photograph of the source and playback monitors and calculating the difference between the displayed timecodes. To minimize local display latency, the monitor was connected directly to the GPU output on both systems.

\section{Results and Analysis}

\subsection{Rate Distortion (RD) Performance}

Table \ref{tab:RDResults} presents the average Peak Signal-to-Noise Ratio (PSNR) and VMAF (4K) scores for configurations with and without Split-Frame Encoding (SFE), with measured PSNR and VMAF (4K) scores averaged across all test sequences and target bitrates. The results indicate that enabling SFE leads to a minor degradation in RD performance in most scenarios.

At 4K resolution, enabling SFE for the H.265/HEVC codec resulted in an average degradation of 0.042 dB in PSNR and 0.053 points in VMAF. For the AV1 codec, the average degradation was 0.040 dB in PSNR and 0.095 points in VMAF. This minor penalty in compression efficiency was most pronounced in the Ultra High-Quality (2 Pass) tuning configuration. In this case, the maximum observed drop for HEVC was 0.410 dB in PSNR and 0.389 points in VMAF, while for AV1, the maximum drop was 0.243 dB and 0.681 points, respectively. For most transcoding applications, the quality degradation from enabling SFE is therefore negligible.

The performance impact of SFE is even less significant at 8K resolution. For the H.265/HEVC codec, enabling SFE resulted in a marginal improvement of 0.002 dB in PSNR and 0.014 points in VMAF. For the AV1 codec, the average degradation was smaller than at 4K, with reductions of 0.006 dB in PSNR and 0.017 points in VMAF, demonstrating near-negligible degradation in RD performance.

\begin{figure*}[t!]
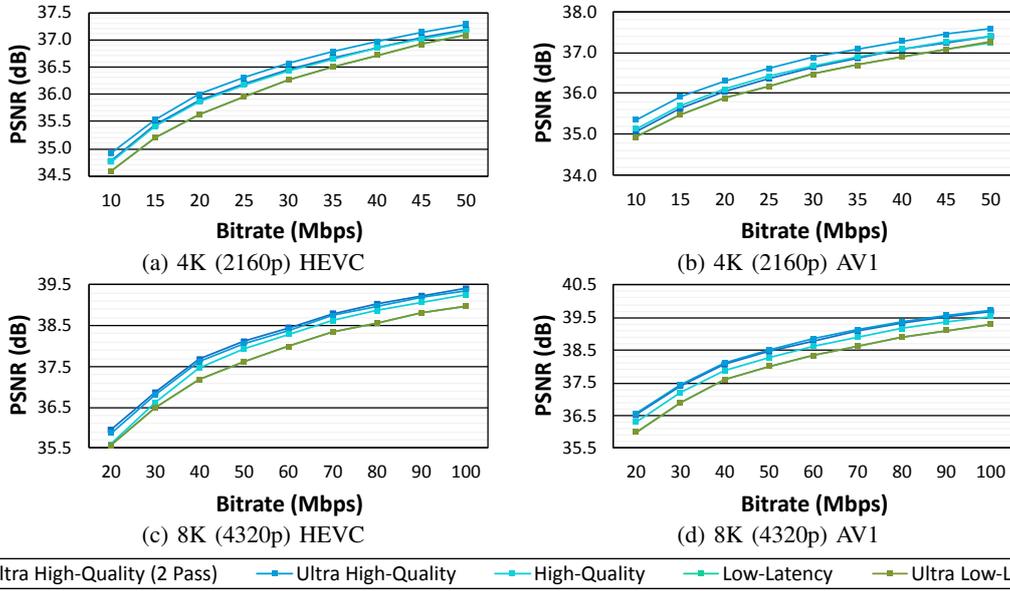

\centering
\begin{subfigure}[t]{.38\linewidth}
  \centering\includesvg[width=0.93\linewidth,inkscapelatex=false]{4K_HEVC.svg}
  \vspace{-1mm}
  \caption{4K (2160p) HEVC}
  \label{fig:4K_HEVC}
\end{subfigure}
\begin{subfigure}[t]{.38\linewidth}
  \centering\includesvg[width=0.93\linewidth,inkscapelatex=false]{4K_AV1.svg}
  \vspace{-1mm}
  \caption{4K (2160p) AV1}
  \label{fig:4K_AV1}
\end{subfigure}\\
\begin{subfigure}[t]{.38\linewidth}
  \centering\includesvg[width=0.93\linewidth,inkscapelatex=false]{8K_HEVC.svg}
  \vspace{-1mm}
  \caption{8K (4320p) HEVC}
  \label{fig:8K_HEVC}
\end{subfigure}
\begin{subfigure}[t]{.38\linewidth}
  \centering\includesvg[width=0.93\linewidth,inkscapelatex=false]{8K_AV1.svg}
  \vspace{-1mm}
  \caption{8K (4320p) AV1}
  \label{fig:8K_AV1}
\end{subfigure}\\
\begin{subfigure}[t]{.95\linewidth}
\vspace{1mm}
  \centering\includesvg[width=0.93\linewidth,inkscapelatex=false]{LegendRDCurve.svg}
  \vspace{-1mm}
\end{subfigure}
\setlength{\belowcaptionskip}{-18pt}

\caption{Rate-Distortion (RD) performance across all tested codecs, resolutions, and tuning configurations at Preset P7 with Split-Frame Encoding (SFE) enabled. Note that the Low-Latency and Ultra Low-Latency curves overlap completely.}
\label{fig:RDCurve}
\vspace{1mm}
\end{figure*}

It was also observed that the Low-Latency and Ultra Low-Latency tuning produced identical encoded outputs with the same PSNR and VMAF scores. The difference between these two tuning modes will be examined in future work. Finally, Fig. \ref{fig:RDCurve} provides the comprehensive RD curves across all tested codecs, resolutions, and configurations at the P7 preset with SFE enabled, serving as a reference for comparative analysis.

\subsection{Encoding Throughput}
\begin{table}[!tbp]
\setstretch{0.83}
\caption{Encoding throughput in frame per seconds (FPS) for each combination of resolution, codec, preset, and tuning. The '\% Diff.' column quantifies the percentage improvement when SFE is enabled.}
\vspace{-1.5mm}
\centering
\label{tab:ThptResults}
\resizebox{8.5cm}{!}{\begin{tabular}{@{}lc ccc ccc@{}}
\toprule 
\multicolumn{8}{c}{UHD 4K (2160p)}\\
\midrule
\multirow{2.5}{*}{Tuning} & \multirow{2.5}{*}{Preset} & \multicolumn{3}{c}{H.265/HEVC} & \multicolumn{3}{c}{AV1}\\ 
\cmidrule(lr){3-5}\cmidrule(lr){6-8} & & No SFE & SFE & \% Diff.& No SFE & SFE & \% Diff.\\
\midrule
\multirow{3}{*}{Ultra High-Quality (2 Pass)}&P7&27.42&22.84&-16.7\%&30.03&23.97&-20.2\%      \\
&P4&67.52&60.61&-10.2\%&48.92&49.55&+1.3\%       \\
&P1&90.42&86.58&-4.2\%&74.63&77.58&+4.0\%        \\
\midrule
\multirow{3}{*}{Ultra High-Quality}&P7&29.40&23.78&-19.1\%&30.36&26.10&-14.0\%      \\
&P4&69.44&61.92&-10.8\%&60.50&59.17&-2.2\%       \\
&P1&95.42&89.05&-6.7\%&89.93&90.42&+0.5\%        \\
\midrule
\multirow{3}{*}{High-Quality}&P7&45.24&88.77&\textbf{+96.2\%}&69.91&135.32&\textbf{+93.6\%}     \\
&P4&129.28&244.80&\textbf{+89.4\%}&108.23&207.04&\textbf{+91.3\%}  \\
&P1&285.31&518.13&\textbf{+81.6\%}&194.17&355.87&\textbf{+83.3\%}  \\
\midrule
\multirow{3}{*}{Low-Latency}&P7&87.99&171.09&\textbf{+94.4\%}&92.42&179.69&\textbf{+94.4\%}    \\
&P4&110.31&213.68&\textbf{+93.7\%}&166.67&314.47&\textbf{+88.7\%}  \\
&P1&285.31&520.83&\textbf{+82.6\%}&281.29&514.14&\textbf{+82.8\%}  \\
\midrule
\multirow{3}{*}{Ultra Low-Latency}&P7&87.99&171.09&\textbf{+94.4\%}&92.55&179.37&\textbf{+93.8\%}    \\
&P4&110.62&212.77&\textbf{+92.3\%}&166.39&316.46&\textbf{+90.2\%}  \\
&P1&285.31&522.19&\textbf{+83.0\%}&281.29&514.14&\textbf{+82.8\%}  \\
\midrule
\multicolumn{8}{c}{UHD 8K (4320p)}\\
\midrule
\multirow{2.5}{*}{Tuning} & \multirow{2.5}{*}{Preset} & \multicolumn{3}{c}{H.265/HEVC} & \multicolumn{3}{c}{AV1}\\ 
\cmidrule(lr){3-5}\cmidrule(lr){6-8} & & No SFE & SFE & \% Diff.& No SFE & SFE & \% Diff.\\
\midrule
\multirow{3}{*}{Ultra High-Quality (2 Pass)}&P7&9.55&11.72&\textbf{+22.6\%}&11.18&13.17&\textbf{+17.8\%}    \\
&P4&22.80&19.40&-14.9\%&14.22&15.35&+7.9\%    \\
&P1&34.22&35.97&\textbf{+5.1\%}&24.93&30.58&\textbf{+22.7\%}    \\
\midrule
\multirow{3}{*}{Ultra High-Quality}&P7&10.90&12.73&\textbf{+16.8\%}&14.83&15.88&\textbf{+7.1\%}    \\
&P4&23.95&19.92&-16.8\%&19.65&18.93&-3.6\%    \\
&P1&37.23&37.71&+1.3\%&33.99&38.76&\textbf{+14.0\%}    \\
\midrule
\multirow{3}{*}{High-Quality}&P7&11.39&22.30&\textbf{+95.8\%}&17.40&33.83&\textbf{+94.5\%}   \\
&P4&32.38&61.65&\textbf{+90.4\%}&26.81&51.39&\textbf{+91.7\%}   \\
&P1&72.67&135.50&\textbf{+86.4\%}&47.66&88.50&\textbf{+85.7\%}  \\
\midrule
\multirow{3}{*}{Low-Latency}&P7&22.20&43.48&\textbf{+95.8\%}&23.25&45.66&\textbf{+96.4\%}   \\
&P4&27.96&54.53&\textbf{+95.0\%}&42.02&81.04&\textbf{+92.9\%}   \\
&P1&72.57&135.50&\textbf{+86.7\%}&71.12&134.41&\textbf{+89.0\%} \\
\midrule
\multirow{3}{*}{Ultra Low-Latency}&P7&22.25&43.44&\textbf{+95.2\%}&23.25&45.66&\textbf{+96.4\%}   \\
&P4&28.00&54.53&\textbf{+94.8\%}&41.98&81.43&\textbf{+94.0\%}   \\
&P1&72.36&136.24&\textbf{+88.3\%}&71.23&132.28&\textbf{+85.7\%} \\
\bottomrule
\end{tabular}}
\vspace{-7mm}
\end{table}

Table \ref{tab:ThptResults} shows the encoding throughput for each combination of resolution, codec, preset, and tuning. For High-Quality, Low-Latency, and Ultra Low-Latency tuning, enabling SFE provided a near-linear improvement in encoding throughput for both 4K and 8K resolutions across the two codecs. At the fastest preset, P1, an improvement of between 81.6\% and 88.3\% was observed, while the slowest preset, P7, displayed the most efficient scaling with an improvement between 93.6\% and 96.4\%. This significant increase in encoding throughput implies that for real-time encoding, enabling SFE can allow for the use of slower presets to achieve better RD performance, offsetting the slight reduction in quality at a given preset.

Conversely, for Ultra High-Quality tuning, enabling SFE at 4K resolution reduced encoding throughput by an average of 11.3\% and 5.1\% for H.265/HEVC and AV1 codecs, respectively. At 8K resolution, the results were mixed, with SFE providing an average improvement of only 2.3\% for H.265/HEVC and 11.0\% for AV1. This gain was observed only at the fastest (P1) and slowest (P7) presets, while the medium-quality P4 preset suffered an average throughput reduction of 6.9\%. 

This suggests a conflict between SFE's parallel architecture and advanced features specific to this preset, which may introduce serial dependencies or require access to global frame statistics that cannot be efficiently shared across the independent encoder, making this combination unsuitable for maximizing performance. Pinpointing the exact mechanism is challenging due to the proprietary, `black-box' nature of NVENC's internal tuning configurations. Therefore, enabling SFE with Ultra High-Quality Tuning is not recommended; other techniques, such as parallel MPEG-DASH segment encoding, should be used to take advantage of multiple NVENC chips instead.

\subsection{Power Consumption}

Understanding the power consumption of the GPU during encoding is essential for designing energy-efficient video encoding systems. While NVENC was fully utilized, the GPU core clock was maintained at 2505 MHz. When encoding to the H.265/HEVC codec, the average power consumption of the GPU board was 38.5 W and 43.0 W with one and two NVENC chips utilized, respectively. For AV1 encoding, these figures increased to 42.0 W and 48.0 W. This indicates that the NVENC hardware consumes slightly more power when encoding to the AV1 codec compared to H.265/HEVC. Notably, these power consumption figures are substantially lower than those of software encoders, which can fully saturate modern microprocessors such as the AMD Ryzen 9 5900X used in this study, resulting in a power consumption of 150 W.

\subsection{End-to-End Latency}

\begin{table}[!tbp]
\setstretch{0.83}
\vspace{1.5mm}
\caption{Measured End-to-End latency in number of frames on 60p Timecode for each combination of resolution, codec, preset, and tuning. `DNR' indicates that the configuration `Did Not Run' or failed to achieve sufficient throughput for real-time encoding.}
\vspace{-1.5mm}
\centering
\label{tab:E2ELatency}
\resizebox{7cm}{!}{\begin{tabular}{@{}lc cc cc@{}}
\toprule 
\multicolumn{6}{c}{UHD 4K (2160p)}\\
\midrule
\multirow{2.5}{*}{Tuning} & \multirow{2.5}{*}{Preset} & \multicolumn{2}{c}{H.265/HEVC} & \multicolumn{2}{c}{AV1}\\ 
\cmidrule(lr){3-4}\cmidrule(lr){5-6} & & No SFE & SFE & No SFE & SFE \\
\midrule
\multirow{3}{*}{High-Quality}&P7&\textcolor{red}{DNR}&5&7&7 \\
&P4&5&5&7&8   \\
&P1&5&5&7&8   \\
\midrule
\multirow{3}{*}{Low-Latency}&P7&5&5&8&8   \\
&P4&5&5&8&8   \\
&P1&5&5&8&7   \\
\midrule
\multirow{3}{*}{Ultra Low-Latency}&P7&5&5&8&8   \\
&P4&5&5&8&8   \\
&P1&5&5&7&7   \\
\midrule
\multicolumn{6}{c}{UHD 8K (4320p)}\\
\midrule
\multirow{2.5}{*}{Tuning} & \multirow{2.5}{*}{Preset} & \multicolumn{2}{c}{H.265/HEVC} & \multicolumn{2}{c}{AV1}\\ 
\cmidrule(lr){3-4}\cmidrule(lr){5-6} & & No SFE & SFE & No SFE & SFE \\
\midrule
\multirow{3}{*}{High-Quality}&P7&\textcolor{red}{DNR}&\textcolor{red}{DNR}&\textcolor{red}{DNR}&\textcolor{red}{DNR} \\
&P4&\textcolor{red}{DNR}&6&\textcolor{red}{DNR}&6     \\
&P1&6&5&7&6         \\
\midrule
\multirow{3}{*}{Low-Latency}&P7&\textcolor{red}{DNR}&\textcolor{red}{DNR}&\textcolor{red}{DNR}&\textcolor{red}{DNR} \\
&P4&\textcolor{red}{DNR}&6&\textcolor{red}{DNR}&6     \\
&P1&6&5&6&6         \\
\midrule
\multirow{3}{*}{Ultra Low-Latency}&P7&\textcolor{red}{DNR}&\textcolor{red}{DNR}&\textcolor{red}{DNR}&\textcolor{red}{DNR} \\
&P4&\textcolor{red}{DNR}&6&\textcolor{red}{DNR}&6     \\
&P1&6&5&6&7         \\
\bottomrule
\end{tabular}}
\vspace{-7mm}
\end{table}

The demand for low-latency live video transcoding has grown with the rise of real-time interactive platforms. It is therefore crucial to understand the latency impact of SFE. Table \ref{tab:E2ELatency} presents the measured End-to-End latency in frames for each combination of resolution, codec, preset, and tuning. The Ultra High-Quality tuning mode was omitted from this evaluation as it is not intended for real-time applications.

At 4K resolution, enabling SFE does not alter the perceived End-to-End latency. It was observed that the AV1 codec adds 2-3 frames of latency, equivalent to 16.7-50.0 ms, compared to H.265/HEVC. Additionally, the Low-Latency and Ultra Low-Latency tuning modes yielded identical latency to the High-Quality tuning, despite disabling B-frame insertion. However, the throughput gain from SFE allows for the use of the P7 preset in real-time H.265/HEVC 4K60p encoding, which is not feasible with a single NVENC chip. As indicated in Table \ref{tab:RDResults}, using the P7 preset with SFE provides a higher quality encoding than using the P4 preset without it, offsetting the RD performance penalty and improving the overall output quality. \looseness=-1

At 8K resolution, enabling SFE with the P1 preset reduces the End-to-End latency by one frame (16.7 ms) for H.265/HEVC, while latency for the AV1 codec remains consistent. This reduction is due to SFE distributing the heavy computational load of 8K video across two encoder chips. This also makes it possible to increase the preset from P1 to P4, yielding an RD performance uplift in the final output compared to using P1 without SFE. Therefore, enabling SFE is proven to be beneficial for real-time 8K encoding scenarios.

\vspace{-0.5mm}
\section{Conclusions and Future Work}

In this paper, the performance of NVIDIA's Split-Frame Encoding (SFE) was evaluated in terms of UHD video transcoding use cases at both 4K and 8K resolutions. The evaluation encompassed Rate-Distortion (RD) performance, encoding throughput, power consumption, and end-to-end latency, comparing SFE-enabled configurations against single NVENC setups across various presets and tuning modes for H.265/HEVC and AV1 codecs.

The results indicate that enabling SFE resulted in a generally negligible degradation in RD performance for most transcoding applications, with a more pronounced impact observed only in the ultra high-quality two-pass configurations. For real-time applications, SFE provides a near-linear improvement in encoding throughput, nearly doubling the frame rate. This enables the use of higher-quality presets for 4K and makes real-time 8K encoding feasible, ultimately improving the final output quality and offsetting the minor RD penalty. However, SFE is not recommended for ultra high-quality offline transcoding, as it can negatively impact throughput. The end-to-end latency analysis demonstrates that SFE does not introduce additional delay at 4K and can reduce latency for 8K encoding. These findings position SFE as a key enabler for high-throughput, real-time UHD transcoding in environments where NVIDIA GPUs are already prevalent, such as data centers and edge servers.

While this study provides a comprehensive analysis of the objective performance of SFE, the impact on subjective quality, including the analysis of potential stitching artifacts at slice boundaries, remains unexplored. This calls for formal subjective evaluations to understand the perceived Quality of Experience (QoE). Furthermore, this analysis was limited by page constraints to PSNR and VMAF; a more detailed Rate-Distortion analysis using metrics like BD-Rate is planned for future work. Moreover, the interaction between SFE and other advanced encoding features, such as adaptive quantization (AQ) and B-frame usage, was not covered in-depth. Subsequent ablation studies will investigate these interactions to identify optimal configurations for various use cases. Finally, we plan to evaluate the performance of SFE in emerging network architectures, such as AI-RAN, to assess its suitability for edge-based live transcoding scenarios.

\vspace{-0.5mm}
\section*{Acknowledgement}

This paper is supported by the Ministry of Internal Affairs and Communications (MIC) Project for Efficient Frequency Utilization Toward Wireless IP Multicasting.




%
\setstretch{1}
\Urlmuskip=0mu plus 1mu\relax
\bibliographystyle{IEEEtran}
\bibliography{b_reference}

\end{document}